# Inverse design of high-$Q$ wave filters in two-dimensional phononic crystals by topology optimization

Hao-Wen Dong[1,2], Yue-Sheng Wang[1,*], and Chuanzeng Zhang[2,*]

[1]Institute of Engineering Mechanics, Beijing Jiaotong University, Beijing 100044, China

[2]Department of Civil Engineering, University of Siegen, D-57068 Siegen, Germany

**Abstract:** Topology optimization of a waveguide-cavity structure in phononic crystals for designing narrow band filters under the given operating frequencies is presented in this paper. We show that it is possible to obtain an ultra-high-$Q$ filter by only optimizing the cavity topology without introducing any other coupling medium. The optimized cavity with highly symmetric resonance can be utilized as the multi-channel filter, raising filter and T-splitter. In addition, most optimized high-$Q$ filters have the Fano resonances near the resonant frequencies. Furthermore, our filter optimization based on the waveguide and cavity, and our simple illustration of a computational approach to wave control in phononic crystals can be extended and applied to design other acoustic devices or even opto-mechanical devices.

**Highlights:**

l  Inverse designs of elastic wave filters in phononic crsytals are presented.
l  Ultra-high-$Q$ filters are realized for the in-plane or out-of-plane elastic waves.
l  Typical Fano resonances are also revealed for most optimized filters.
l  Optimized cavities can be the ideal components for the other acoustic devices.

## I. Introduction

As the acoustic or elastic analogue of photonic crystals (PtCs), phononic crystals (PnCs) [1] can be employed for controlling the wave propagation behaviors and opening novel routes for new types of highly effective acoustic or elastic wave devices. Due to the possibility of prohibiting wave propagation within certain frequency bands, PnCs combined with appropriate line or point defects can be utilized to generate the waveguides [2-5], cavities [6, 7] and filters [5, 8, 9, 10], etc. Thanks to the potential to achieve a large quality ($Q$) factor and low-loss resonators in ultra-compact cavities, PnC devices are particularly suitable for a wide range of applications from the acoustic frequency communication to the medical ultrasound. In particular, wave filters in PnCs have become the compelling passive devices in wave signal processing systems. By introducing the waveguides and cavities, one can use the wave filters to manipulate the frequencies, improve the frequency selectivity (wavelength resolution), reduce the insertion loss, and decouple the high-$Q$ resonator. In addition, acoustic wave filters can be extended to the frequency division multiplexing in acoustic systems and electro-acoustic transducer design. Inspired by the similar selection and transference property of PtC filters, several PnC filters have been demonstrated in different dimensions and systems. For one-dimensional (1D) PnCs, Zhang et. al [10] showed that the broadband wave filtering can be achieved in the PnCs with the hierarchical structure. Besides, the broad passband and narrow passband filters were realized by the impedance-mirroring [11]. And for the two-dimensional (2D) case, the tunable narrow passband filters [5, 8] and channel drop filters [9] were proposed by strategically placing defects in solid/water PnCs. By adjusting the geometrical and material parameters, the corresponding active filtering and coupling can be improved or tuned. However, the proposed interesting PnC filters so far have some shortcomings. The filters in Refs. [5, 8, 9] were designed in the water matrix, where the energy loss is larger than that in solid materials. In order to ensure the coupling, the defects usually involve inclusions of different materials [8], which is a challenge for the fabrication and application. Besides, the significant deficiency is that the obtained filters have a relatively small $Q$ factor. Since $Q$ determines the resolution of the filtering operation, a large $Q$ factor will greatly improve the overall performance for the application. Furthermore, a large $Q$ factor will reduce the power threshold in acoustic devices. Hence, it is necessary to design high-$Q$ solid PnC filters at given operating frequencies by using simple coupling structures.

* Corresponding author: yswang@bjtu.edu.cn
* Corresponding author: c.zhang@uni-siegen.de

To design the engineering structures with a high performance, topology optimization has been widely used in various fields. As a free-form design method, topology optimization offers more freedom and new routes to explore the beneficial structures. In the last decade, with the help of the topology optimization, many high-performance structures and devices in PtCs [12, 13], PnCs [14-18] and phoxonic crystals (PxCs) [19] have been successfully designed and proposed. In particular, compared with the hand-designed solutions, the topology optimized results can provide a broader guidance for applications. Fortunately, the rapid development of novel fabrication methods and 3D printing will be helpful to realize the optimized structures.

In view of the necessity of high-$Q$ filters and the advantages of topology optimization, topology optimization of 2D PnC filters in a single solid material is presented in this paper. To this end, the finite element method (FEM) is applied to calculate the dispersion relations and wave transmission properties. The genetic algorithm (GA) [20] is used as the inverse method for the present single-objective optimization. For incident shear horizontal (SH) or longitudinal (P) waves at different operating frequencies, various high-$Q$ filters by only optimizing the cavity topologies are obtained. We also show the effect of the distance between the cavity and the waveguide on the optimization and performance of the optimized filters. Based on the optimized cavity, we demonstrate different novel applications in PnC devices. The present paper is organized as follows. First, the formulation of the topology optimization problem is described in Sec. II. Then, the optimized structures and the discussions are presented in Secs. III and IV. Finally, some conclusions will be drawn in Sec. V.

## II. Problem formulation

Figure 1 presents the schematic sketch of the optimization problem for a 2D PnC drop filter based on a waveguide and a single cavity. In order to trap and emit waves at a given frequency, the cavity is taken as the design domain for the topology optimization as shown in Fig. 1(c). To reduce the computational effort, the cavity is assumed to have a square symmetry, i.e., invariant under the mirror reflection with respect to the $x$-$z$-plane and $y$-$z$-plane and under a 90° rotation around the $z$-axis. When the resonant frequency matches the frequency of the incident wave along the waveguide, it is possible to transfer a part of the energy into the cavity and then reduce the output energy. In principle, a stronger coupling efficiency allows a better drop filtering. The fixed single line defect (W1) waveguide in Fig. 1(a) guides the incident P or SH waves emitted from the left port of the waveguide. The computational model in Fig. 1(a) is terminated with the non-reflecting boundaries. Fig. 1(b) shows the dimensions of the unit-cell and the waveguide.

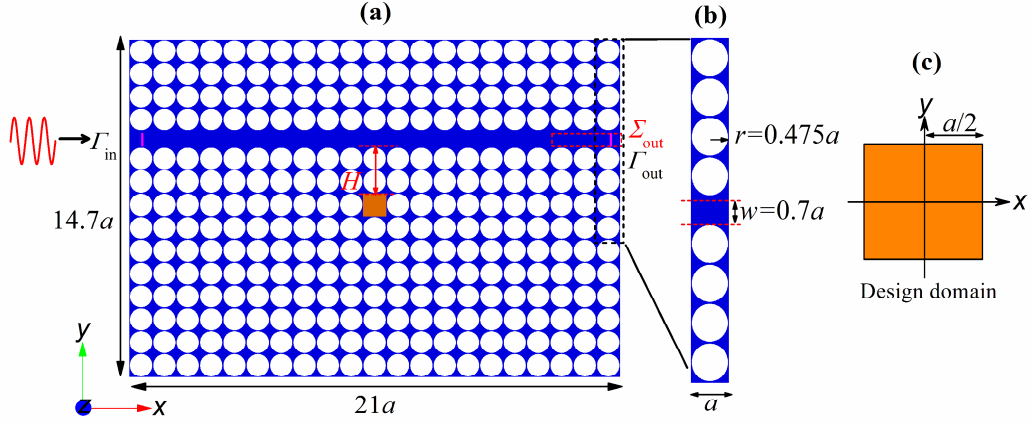

Fig. 1. A PnC drop filter with a W1 waveguide and a single defect cavity: **(a)** Schematic sketch of a $21a \times 14.7a$ structure having a square lattice of vacuum with a lattice constant $a$. The point defect area is the design domain of the optimization, which considers the incident P or SH waves from the left port of the waveguide. The energy flux in the area of $\Sigma_{out}$ is computed during the optimization. **(b)** Dimensions of the unit-cell and the waveguide. The radius of the circular vacuum holes is $r$ and the width of the waveguide is $w$. **(c)** Design domain with the assumed square symmetry. Only one-eighth of the unit-cell needs to be considered in the optimization.

We assume that a plane time-harmonic elastic wave propagates in the structure. The equations of wave motion in the elastic solids are given as [21]

$$C_{ijkl} u_{k,lj} + \rho \omega^2 u_i = 0, \tag{1}$$

where $i, j, k, l = 1, 2, 3$; $C_{ijkl}$ represents the elasticity tensor; $\rho$ declares the mass density; and $u_i$ denotes the displacement vector. Here and throughout the paper, the common time-harmonic factor $e^{i\omega t}$ (where $\omega$ is the circular frequency) is omitted for the sake of brevity.

To evaluate the transmitted energy across an interface, the instantaneous Poynting vector [21] is used, which is defined by

$$P_j(\mathbf{x}, t) = -\sigma_{ij}(\mathbf{x}) e^{i\omega t} \frac{\partial u_i}{\partial t}, \tag{2}$$

in which $\sigma_{ij}$ is the components of the stress tensor. We compute the normalized transmission $T$ of the filter along the $x$ direction defined as

$$T = \frac{\oint_{\Gamma_{out}} \int_0^{2\pi/\omega} P_x \, dt \, d\Gamma}{\oint_{\Gamma_{in}} \int_0^{2\pi/\omega} P_x \, dt \, d\Gamma}, \tag{3}$$

where $\Gamma_{in}$ and $\Gamma_{out}$ are the lines in the waveguide shown in Fig. 1, and $P_x$ is the component of the instantaneous Poynting vector along the $x$-axis.

The time-harmonic displacement solution is denoted by $\mathbf{U} = \mathbf{u}(\mathbf{x})e^{i\omega t}$. Then, for either in-plane (P) or out-of-plane (SH) waves, we can obtain the following system of linear algebraic equation from the FEM discretization of Eq. (1) as

$$(\mathbf{K} - \omega^2 \mathbf{M})\mathbf{U} = \mathbf{f}, \tag{4}$$

where $\mathbf{K}$ is the stiffness matrix; $\mathbf{M}$ denotes the mass matrix; and $\mathbf{f}$ represents the incident time-harmonic force vector. The displacement, stress and velocity fields for different incident elastic waves are computed by using the commercial FEM software ABAQUS [18, 19].

To obtain the drop filtering at a given frequency for the concerned spectral range, the total displacement field in the area of $\Sigma_{out}$ in Fig. 1(a) must be minimized. The selection of the energy instead of the normalized transmitted displacements aims at increasing the objective sensitivity of the cavity and accelerating the optimization procedure. Otherwise, the topology optimization cannot find the evolution direction. Hence, the objective function for minimizing the averaged transmitted power in the outport $\Sigma_{out}$ of the waveguide is defined as:

$$\text{Minimize：} \quad E(\omega) = \frac{\omega}{2\pi S} \oint_{\Sigma_{out}} \int_0^{2\pi/\omega} P_x \, dt \, dS, \tag{5}$$

$$\text{Subject to：} \quad \min_\phi (e) \geq e^*, \tag{6}$$

$$\rho_i = 0 \text{ or } 1 \quad (i=1,2,\cdots N \times N), \tag{7}$$

where $E$ is the integrated energy intensity, $S$ is the area of outport $\Sigma_{out}$, and $\phi$ represents the topological distribution within the cavity. We characterize the design domain as a binary matrix, like in Ref. [19]. In Eq. (7), $\rho_i$ denotes the material density of the $i$th element (the cavity is meshed by $N \times N$ elements) and declares the absence (0) or presence (1) of the solid. In particular, we introduce a geometrical constraint in Eq. (5) to overcome the typical mesh-dependence problem [22, 23] in the optimization and to make an easy fabrication of the optimized structure. It implies that the minimal width $e$ of the solid connections appearing in the cavity should be larger than the prescribed limited value $e^*$ which is selected as $a/30$ in this paper. For more details on this constraint, we refer to the Refs. [19, 23].

We use the single-objective genetic algorithm (GA) to solve the optimization problem described by Eqs. (5)-(7). As an evolutionary algorithm, GA applies the principles of natural evolution to optimize a given objective. In the GA optimization, different solutions to the problem represent different generated structures. In other words, a cavity with the solid-vacuum distribution is defined as an individual evolved in the optimization. The details of the optimization procedure can be stated as follows:

(1) Start with an initial random population which has a number of individuals.
(2) For each structure, we use the "abuttal entropy filter" method [23] to filter its topology to some extent. Some isolated elements are removed to eliminate the mesh-dependency, and some isolated voids are filled up to improve the structural strength.
(3) Wave propagation in every PnC drop filter is simulated by ABAQUS. Then, the fitness function value based on Eqs. (5) and (6) is assigned to each individual.
(4) A set of operations, namely, the reproduction (selection), crossover and mutation operators, are performed for every individual to create a new offspring generation. For the reproduction, some individuals will be copied to breed a new generation according to the fitness.
(5) The crossover operator is applied for the selected parent population. Exchanging the partial genes of the parent individuals will produce two new offspring individuals.
(6) An adaptive mutation operator [18] is applied for every individual. Then, the final new population is generated. To accelerate the optimization procedure, the elitism strategy is adopted. The best individual in the current generation is copied from generation to generation.
(7) If a fixed number of generations are completed, the final optimized filter and cavity are generated. Otherwise, return to step (2). As this process evolves, better and better solutions can be found.
(8) If necessary, map the final optimized structures into a refined mesh and run a new round of the optimization.

More details on the above mentioned procedure can be found in our previous works [18, 23].

## III. Filter optimization and filtering properties

We consider a waveguide-cavity structure based on a square-latticed porous PnC with the solid material of the mass density $\rho$=2330 kgm$^{-3}$, Lamé's constant $\lambda$=85.502 GPa and shear modulus $\mu$=72.835 GPa [16, 23]. For the objective function calculation, the commercial FEM software ABAQUS is applied. The used parameters of the GA are the population size $N_p$=20, the crossover probability $P_c$=0.9, and the mutate probability $P_m$=0.02. To achieve sufficiently smooth edges with a low computational cost, the two-step optimization technique is adopted. First, the design domain is meshed into 30×30 pixels (elements). Through an evolution of 1000 generations the maximum of the fit becomes steady-state. Then, the near-optimal structure is mapped into 60×60 pixels and used as the seed solution of the new run of the optimization procedure. After 2000 generations, the final optimized cavity is obtained. All optimizations were accomplished in about 86 h on a Linux cluster with 16 cores of Intel Xeon E5-2660 at 2.20 GHz. Below we present some representative results to illustrate various interesting aspects of the solutions for the 2D PnC drop filter optimization problem.

### (A) Out-of-plane SH wave filtering

Before presenting the obtained optimization results, we briefly discuss the band structure and the guided modes of the waveguide shown in Fig. 1(b). Figures 2(a) and 2(b) show the results of the out-of-plane and the in-plane wave modes, respectively. Here the normalized frequency $W=\omega a/\pi c_t$ is introduced for convenience, where $c_t$ is the transverse wave velocity. The relative bandgap widths of the two wave modes are 77.2% (0.346-0.781) and 44.4% (0.392-0.616), respectively. We can observe three modes existing in the bandgap for the out-of-plane and in-plane waves in Figs. 2(a) and 2(b). Of course, one should use the single mode for the filter optimization. In fact, the variation of the slope can cause the group-velocity dispersion [24]. Therefore, it is usually desirable to operate where the slope is nearly constant for decreasing the losses and minimizing the distortion of the signal. However, to show the robustness and effectiveness of the present topology optimization design, we select different frequencies with the different velocities or modes to design the drop filter. The selected frequencies which will be used in the following optimization are marked in Fig. 2 by the hollow circles. For the frequencies of the in-plane waves, the modes are mainly dominated by the longitudinal vibrations in spite of the existence of the shear motions. It is noted that the five modes shown in Fig. 2 are either symmetric or antisymmetric.

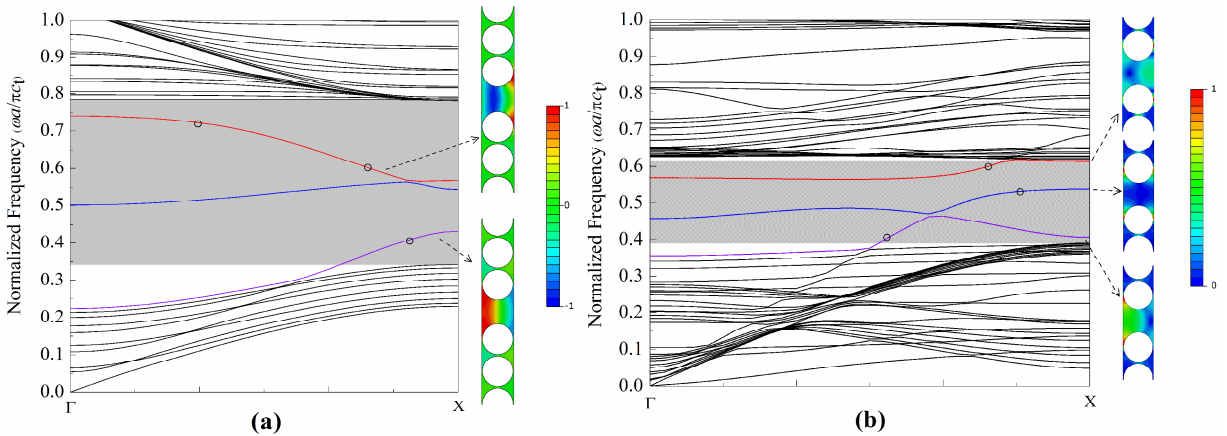

Fig. 2. The dispersion curves for the guided wave modes (solid lines in the grey areas) of a PnC waveguide. (a) Out-of-plane wave mode, (b) In-plane wave mode.

We first show the optimization results for the incident SH wave at three different normalized frequencies $38254.7 \times 10^{-5}$, $62631.42 \times 10^{-5}$ and $71265.12 \times 10^{-5}$ (see Fig. 2(a)). The minimal effective value of the eigen-frequency is about 1 Hz with its corresponding normalized value being $3.58 \times 10^{-6}$ which is within the computational accuracy. Note that the minimal effective value of the normalized frequency in this paper is set as $3.58 \times 10^{-6}$. The corresponding optimized cavity topologies, the normalized transmission and the spatial distribution of the displacement field for the filter are shown in Fig. 3. Overall, the transmission performance shows that the drop filtering properties are obtained for different frequencies as wished. After optimizations, the output energies are reduced to nearly zero at the target frequencies. The drop efficiencies are 99.5% [Fig. 3(a)], 97.8% [Fig. 3(b)] and 99.4% [Fig. 3(c)], respectively. The filtering performance at $Rf=38254.7 \times 10^{-5}$ in Fig. 3(a) is shown in Fig. 3(d). The wave couples into the cavity, preventing the wave propagation in the waveguide. Compared with the lower and upper frequencies, the optimization for the middle frequency in Fig. 3(b) can achieve a $Q$ factor as large as 107528. This verifies the fact that the slope at $62631.42 \times 10^{-5}$ in Fig. 2(a) is nearly constant. Therefore, the smaller energy loss results in the larger $Q$ factor. In addition, three different cavities are optimized at three different frequencies. We can find that the cavities in Figs. 3(b) and 3(c) have the similar geometrical feature, namely, central hole with four symmetric vacuum slots. This is attributed to the same guided modes (see Fig. 2(a)) for these two frequencies, resulting in the similar coupling performance. Of course, this also proves the validity of our optimization results. Besides, we note that a certain bandwidth of the filtering appears around two resonant frequencies which are marked as $R_a$ and $R_b$ in Figs. 3(c). Their corresponding displacement fields $u_z$ of the cavity are presented in Fig. 3(e). By looking inside the cavity, two different vibrations can be found at $R_A$ and $R_B$, respectively. However, both cavity resonances can drop the incident wave from the oblique direction. For the neighboring frequencies, the similar drop performance can be activated as well.

Particularly, it is noteworthy that the asymmetric transmission is obtained for the optimized cavity as shown in Fig. 3(b). Because most optimized filters with a large $Q$ factor in this paper have asymmetric transmissions, this property will be discussed in details in Sec. III(C).

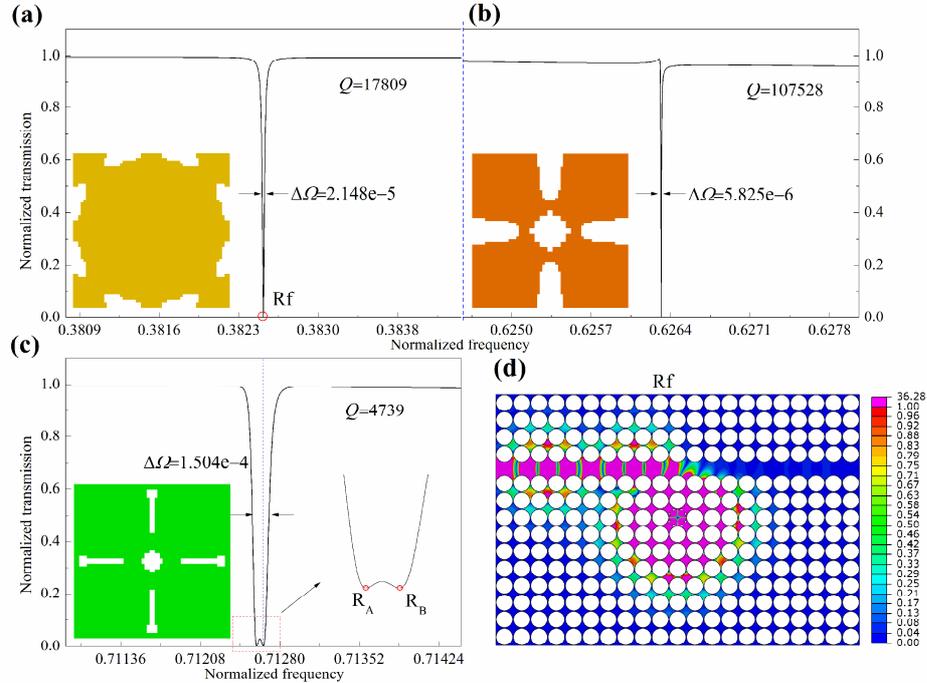

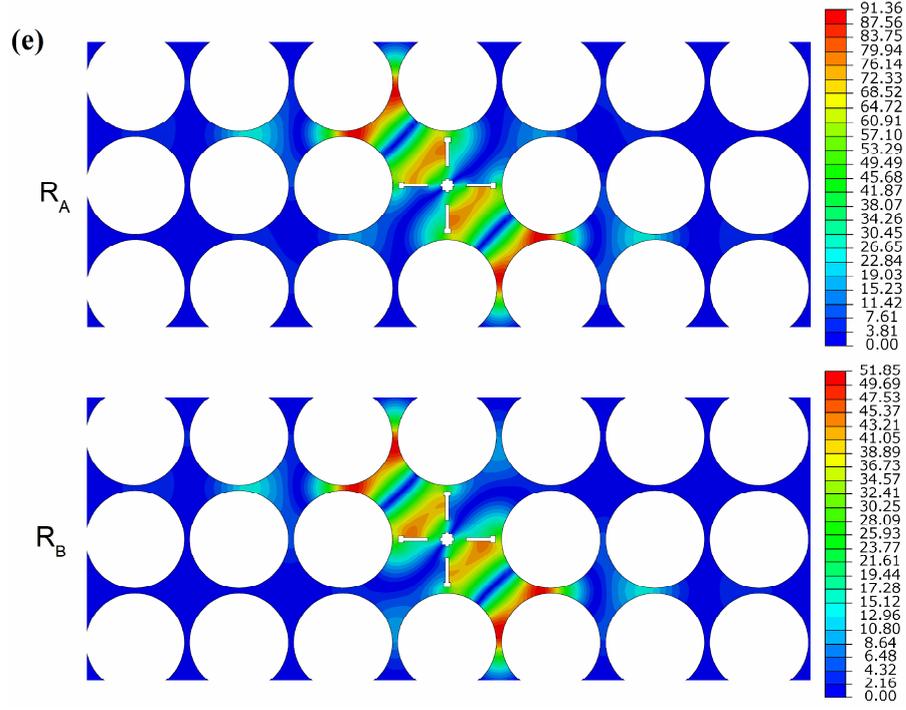

Fig. 3. Optimized cavities and their corresponding transmission spectra with $H=2a$ for the incident SH wave at the normalized frequencies of $38254.7 \times 10^{-5}$ **(a)**, $62631.42 \times 10^{-5}$ **(b)** and $71265.12 \times 10^{-5}$ **(c)**. **(d)** The spatial distribution of the amplitude of the displacement field $u_z$ at the resonant frequency Rf for the structure in (a). **(e)** Responses inside the cavity at two frequencies $R_A$ and $R_B$ marked in (c). Unless otherwise specified, the size of the target area $\Sigma_{out}$ in our optimization is selected as $3a$.

To show the effect of the model size on the optimization, the optimization results for the smaller ($19a \times 12.7a$) and larger ($23a \times 16.7a$) filters are illustrated in Fig. 4. The operating frequency of the incident SH wave is selected as the same as that in Fig. 3(b). It is observed that two optimized cavities have the same topological feature as that in Fig. 3(b). Moreover, the optimized cavity of the smaller filter is identical with that in Fig. 3(b). We can find that the larger filter is able to obtain a larger $Q$ factor. A smaller filter will result in a smaller $Q$ factor. This can be easily understood by considering the fact that a larger model will reduce the energy loss. Besides, Fig. 4 demonstrates that the optimized filters with the size of $21a \times 14.7a$ used in this paper will keep a steady-state filtering property if we appropriately increase or decrease the size of the optimized structures. However, the change of the model size will result in a larger or a smaller $Q$ factor.

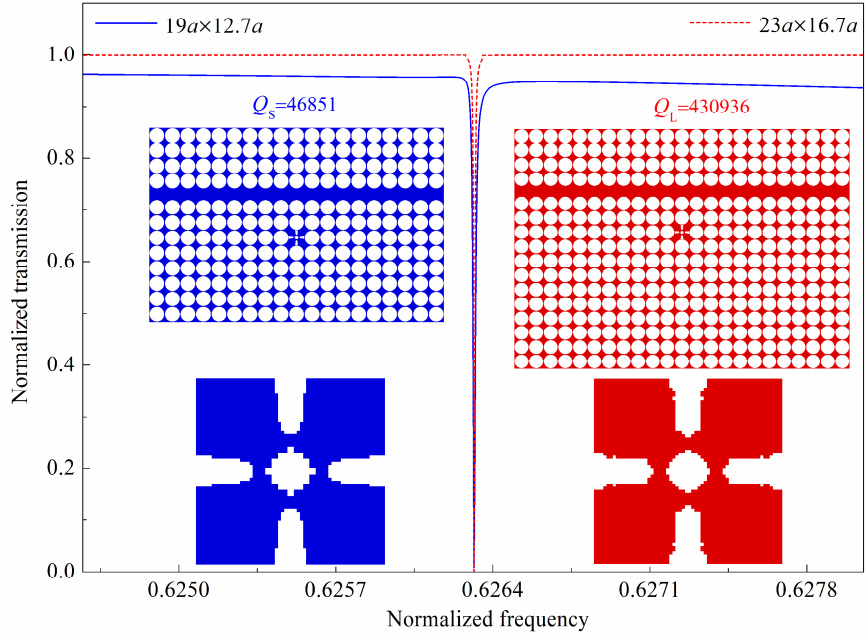

Fig. 4 Optimization results for the smaller ($19a\times12.7a$) and larger ($23a\times16.7a$) filters. As in Fig. 3(b), the incident SH wave at the normalized frequency of $62631.42\times10^{-5}$ is considered. The two optimized cavities and $Q$ factors of the smaller and larger models are also shown.

As mentioned in Sec. II, the target area $\varSigma_{out}$ is selected as $3a$ to accelerate and encourage the optimization. To explore its effect on the optimization, we replace the target area $\varSigma_{out}$ by $2a$ or $4a$ and present the corresponding optimized results in Fig. 5. For comparison, the operating frequency is selected as the same as that in Fig. 3(b). Obviously, a small target energy area $2a$ can hardly obtain the filtering property at the prescribed frequency. In contrast, the larger areas of $3a$ in Fig. 3(b) and $4a$ in Fig. 5 can effectively provide the ideal filtering properties. In fact, the optimized filter with $\varSigma_{out}=4a$ has the same geometry as that in Fig. 3(b). Thus, we generally suggest to use a large outport area for the filter optimization.

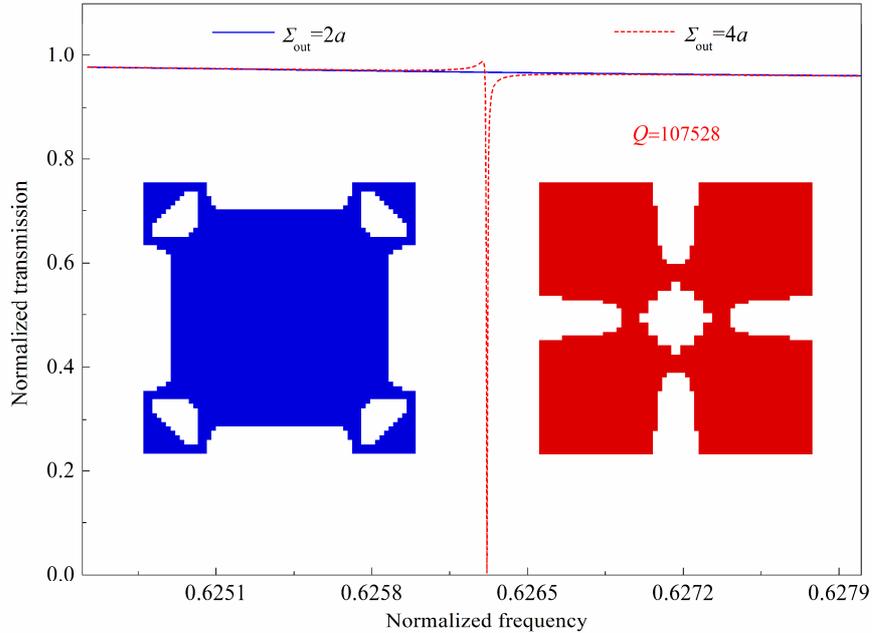

Fig. 5 Transmission spectra of the optimized filters with different target areas $\varSigma_{out}=2a$ and $4a$ for the optimization. As in Fig. 3(b), the incident SH wave at the normalized frequency of $62631.42\times10^{-5}$ is considered. The two optimized cavities and $Q$ factors are also shown.

In order to analyze the drop filtering of the optimized cavity in Fig. 3(b), we examine the corresponding localized cavity modes whose intensity spectrum and displacement filed are plotted in Fig. 6(a). The full width of the spectrum at the half maximum and the $Q$ factor are about $1.362 \times 10^{-6}$ and 459848, respectively. Obviously, the cavity mode has a high symmetry. The vibration mainly occurs in the middle region and the narrow connections between the vacuum holes. Figure 6(b) shows the filtering response and the vibration inside the cavity at the optimized frequency in Fig. 3(b). It is seen that the cavity and the junction of the waveguide and the defect exhibit stronger vibration than other regions. When the incident wave propagates to the junction region, the cavity resonance will be excited by the vibrations of the narrow connections, resulting in the strong coupling. At the resonance, the defect has a symmetric vibration which is parallel to the waveguide. However, the strongest vibration is localized in the middle area. This is obviously distinct from the vibrations in Fig. 3(e). Therefore, the discrepancy between these two systems in the $Q$ factor should be attributed to the symmetry of the cavity mode and the degree of the vibration localization.

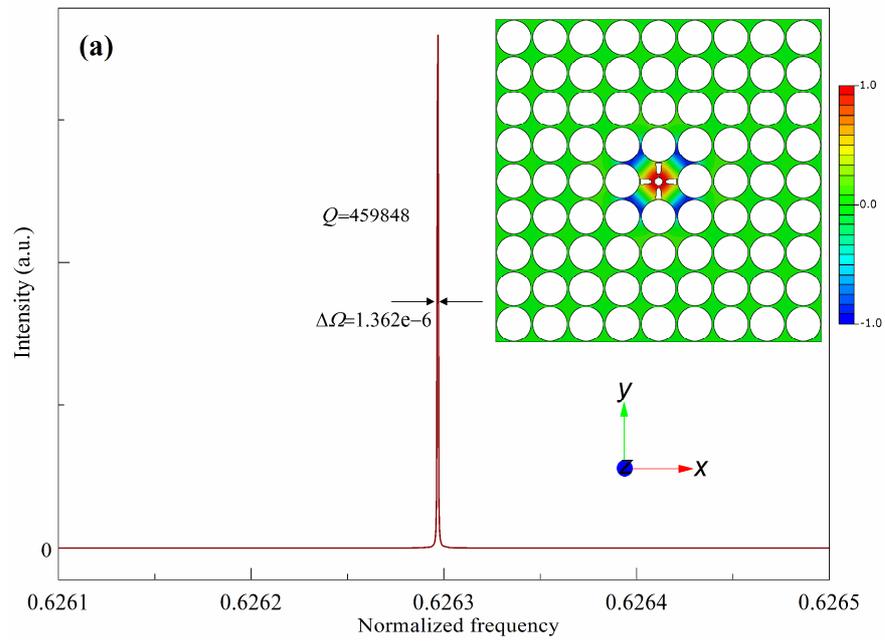

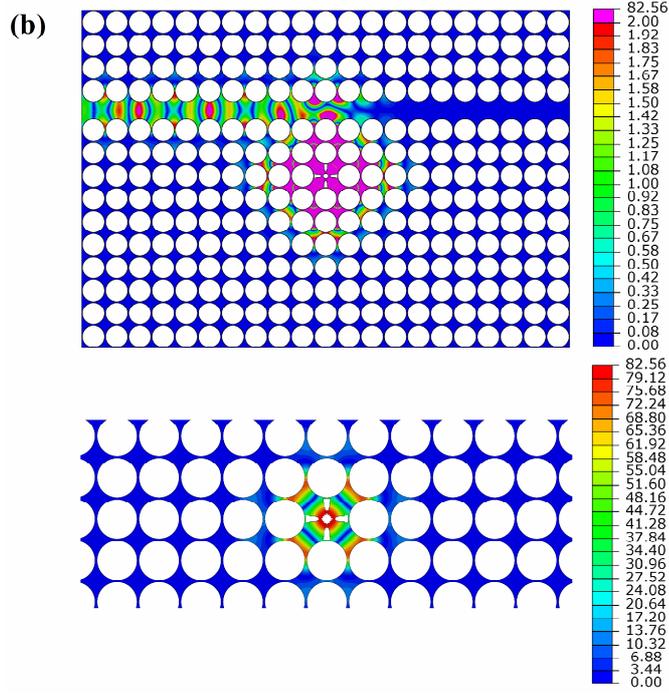

Fig. 6. **(a)** Intensity spectrum of the transmitted wave in the optimized cavity in Fig. 3(b). The inset shows the spatial distribution of the amplitude of the displacement field $u_z$ of the cavity mode. **(b)** The filtering response (upper) and vibration inside the defect cavity (lower) at the resonant frequency in Fig. 3(b).

To investigate the influence of the distance $H$ between the cavity and the waveguide, we perform the optimization for different values of $H$ at the frequency of $62631.42 \times 10^{-5}$, see Fig. 7. Obviously, the distance $2a$ provides an optimized filter with the highest $Q$ factor. When the cavity is close to the waveguide, both the $Q$ factor and the drop efficiency of the optimized structure will become lower. The optimized filter with $H=a$ has nearly a 100% drop filtering over the entire spectrum, except at the resonant frequency. The inset in Fig. 7 shows the corresponding drop filtering. By comparing the responses in Figs. 7 and 6(b), we can find that more energy in Fig. 7 is feeding back into the waveguide when the resonance is excited, leading to the decrease of the $Q$ factor. However, the drop filtering will not even be optimized if the cavity is far from the waveguide, for instance in the case of $H=3a$. Besides, the optimized cavities show that a larger distance $H$ will yield a simpler structure in the optimization. This is due to the fact that for a small $H$ the cavity can enhance the coupling with the waveguide and reduce the energy loss into the waveguide simultaneously, while for a large $H$ the focus turns to perfectly match its resonant frequency with the target frequency.

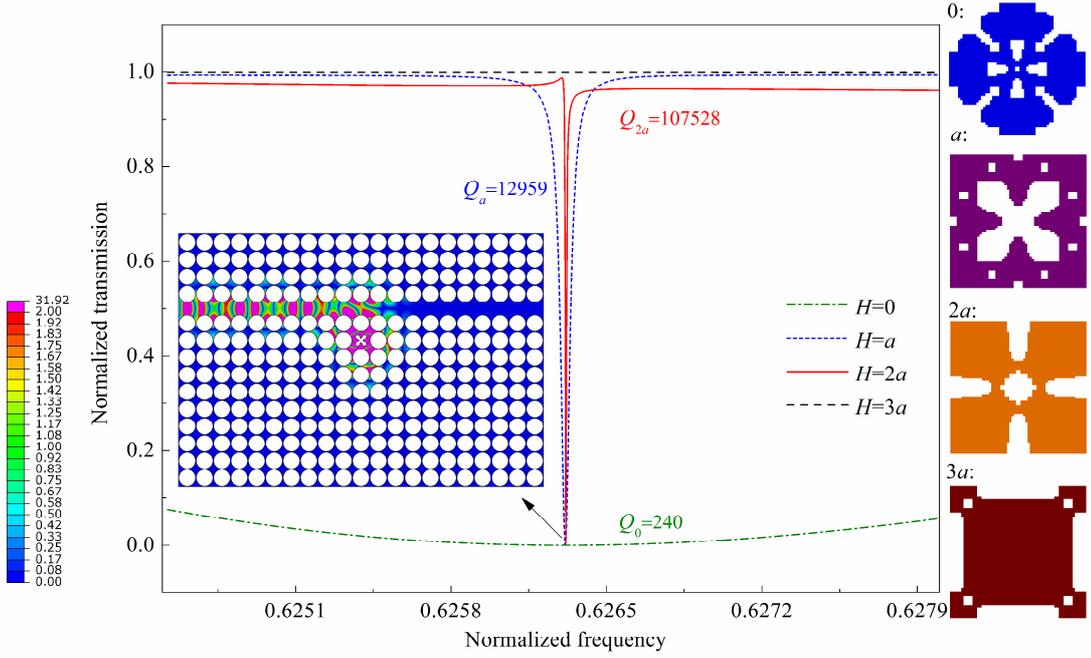

Fig. 7. Optimization results at the filtering frequency of 62631.42x10$^{-5}$ for $H=$ 0, $a$, 2$a$ and 3$a$. The four right figures are the optimized cavities. The inset shows the spatial distribution of the amplitude of the displacement field $u_z$ for the drop filtering with $H=a$.

Motivated by the above analysis, we move up or down the optimized cavity with $H=2a$ to demonstrate its robust filtering property, see Fig. 8. Obviously, both structures with $H=a$ and $H=3a$ can still exhibit a drop filtering. However, compared to the result with $H=2a$, their filtering frequencies change with different degrees. The resonant frequencies for $H=a$ and $3a$ are 62681.5x10$^{-5}$ and 62629.74x10$^{-5}$, respectively. In particular, the frequency of the structure with $H=2a$ is 0.0027% larger than that with $H=3a$ whose $Q$ factor is as large as 4344376. Combining the results for $H=3a$ in Figs. 7 and 8, the small difference of 0.0027% in the resonant frequencies demonstrates the accuracy of our optimized results. Indeed, the filter design is very sensitive to any small change of the target operating frequency. Besides, the filtering responses in Figs. 6(b) and 8 show that the better matched mode is responsible for a larger $Q$ factor.

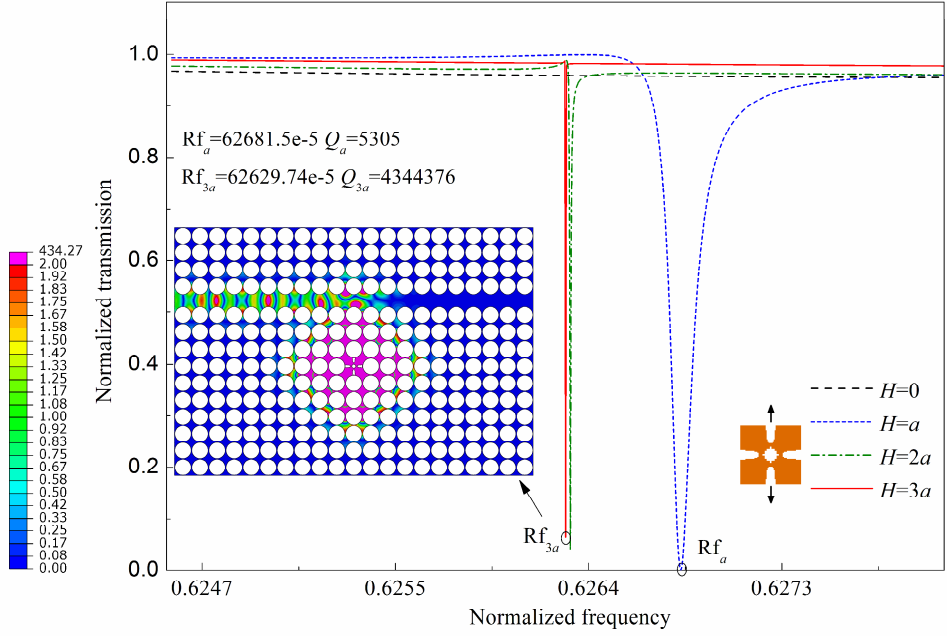

Fig. 8. Transmission spectra of the structures by moving up ($H=0$ or $a$) or down ($H=3a$) the optimized cavity in Fig. 3(b) from the location of $H=2a$. The inset shows the spatial distribution of the amplitude of the displacement field $u_z$ at the resonant frequency.

**(B) In-plane P wave filtering**

We now consider the optimization of elastic wave filters for an incident P wave at three different frequencies of $39483.1 \times 10^{-5}$, $52430.17 \times 10^{-5}$ and $61029.69 \times 10^{-5}$ (see Fig. 2(b)). The results are illustrated in Fig. 9. Three optimized structures show a drop filtering property at the target frequencies. However, the structures are more complex than those for an incident SH wave as shown in Fig. 3. This is due to the coupling between the longitudinal and the transverse wave modes. Compared with the frequencies near the lower and upper bandgap edges, the optimized structure for the frequency $52430.17 \times 10^{-5}$ has a better drop filtering capacity and a large $Q$ factor of 123381. Its filtering performance is shown in Fig. 9(d). Like the results in Fig. 6, the energy loss from the cavity determines the $Q$ factor. The four exterior parts of the structure are simulated by the infinite elements (CINPE4 in ABAQUS/Standard) to prevent the reflection from the boundaries. A large vibration in the waveguide occurs in the narrow connections. Then, the cavity mode is excited to drop the energy into the cavity. Therefore, the optimized cavity mode is able to effectively filter the complex in-plane waves.

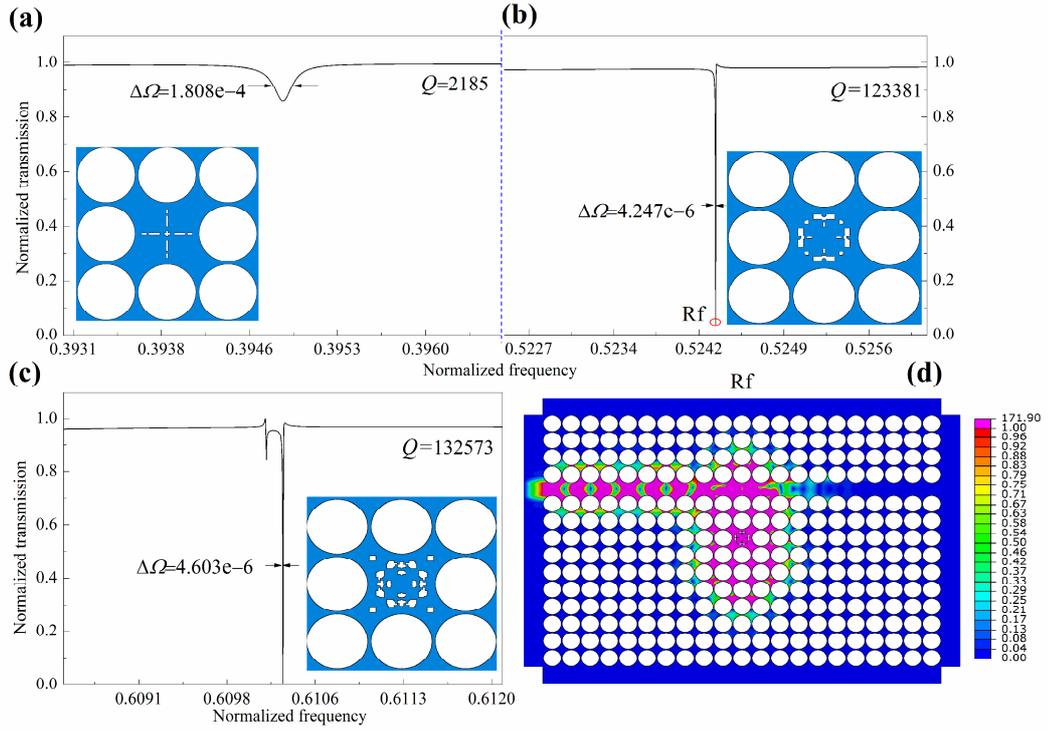

Fig. 9. Optimized cavities and their corresponding transmission spectra of the optimized structures with *H*=2a for the incident P wave at the normalized frequencies 39483.1x10$^{-5}$ **(a)**, 52430.17x10$^{-5}$ **(b)** and 61029.69x10$^{-5}$ **(c)**. **(d)** The spatial distribution of the amplitude of the displacement field $u = \sqrt{u_x^2 + u_y^2}$ at the resonant frequency Rf for the structure in (b).

To further show the effect of the operating frequency on the optimized filters, we perform the optimization for the P wave with 1% increase or 1% decrease of the target frequency in Fig. 9(b), see Figs. 10(a) and 10(b). It can be concluded here that different frequencies lead to different cavities, and a higher frequency yields a larger *Q* factor. So, the optimized filters are very sensitive to the operating frequency. Using the topology optimization, it is able to obtain the simple optimized cavities and high-*Q* filters for even very close target frequencies.

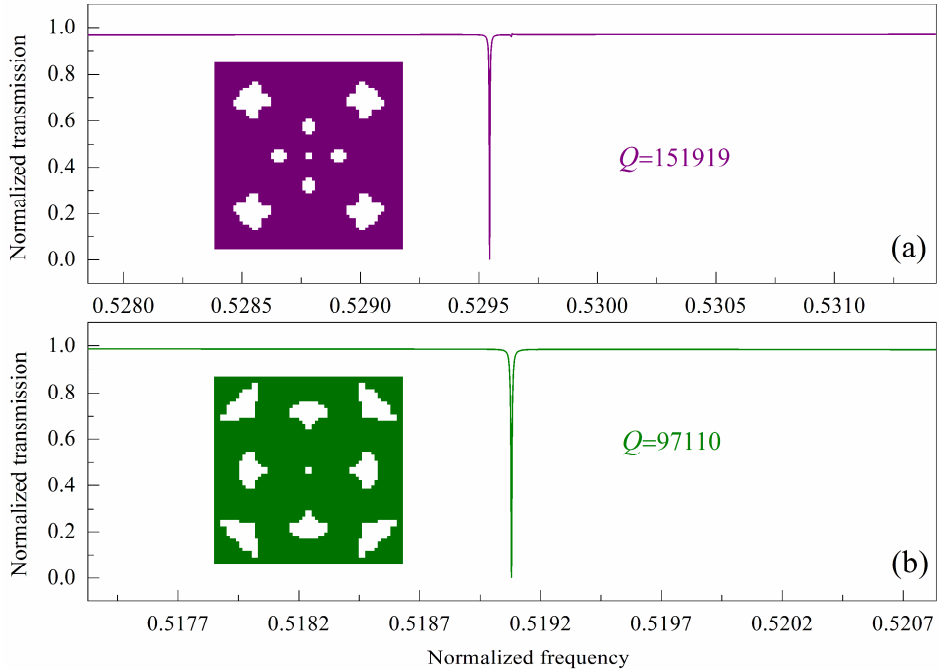

FIG. 10. Optimization results for the incident P wave at the normalized frequencies 52954.48x10$^{-5}$ (**a**) and 51905.73x10$^{-5}$ (**b**) which are 1% larger and 1% smaller than 52430.17x10$^{-5}$ in Fig. 9(b), respectively. The two optimized cavities and *Q* factors are also shown.

To seek a simple filter for the incident P wave, we replace the based unit-cell in Fig. 9 by a local resonant structure whose geometrical parameters are displayed in Fig. 11. The same distance *H*=2*a* between the cavity and the waveguide is selected because this distance can yield a good performance for both SH and P waves. The width of the waveguide is 0.7*a*. Like the unit-cell, the optimized cavity in Fig. 11 has four solid lumps in the center, showing the possibility of the local vibration response. For the target frequency of 59768.96x10$^{-5}$ (point R in Fig. 11), the optimized filter can achieve an ultra-high *Q* factor of 1562581 and a high filtering efficiency of 86%.

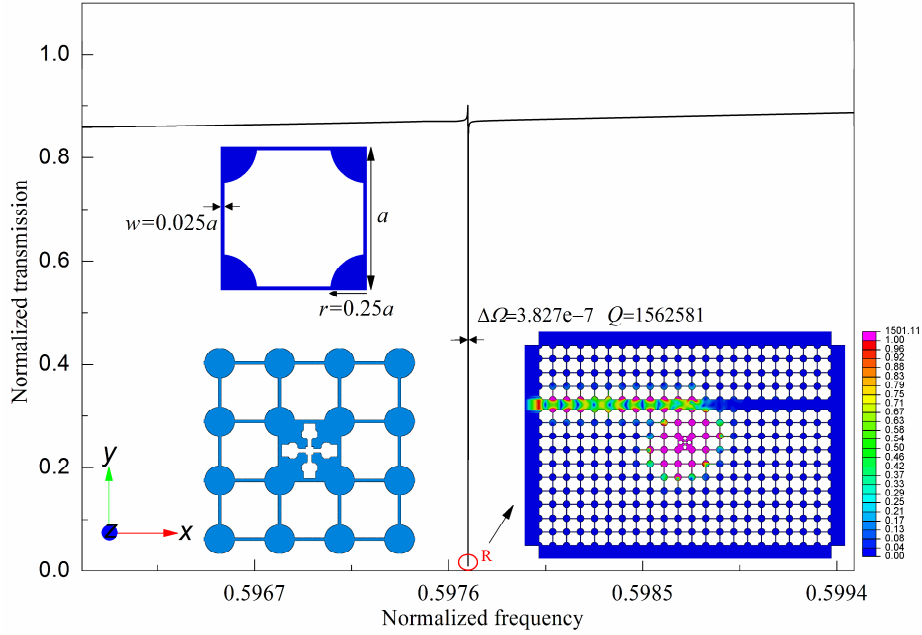

Fig. 11. Transmission spectrum of the optimized filter for the incident P wave using the changed unit-cell structure. The target optimization frequency is 59768.95x10$^{-5}$. The insets show the unit-cell structure, optimized cavity and spatial distribution of the amplitude of the displacement field $u = \sqrt{u_x^2 + u_y^2}$ at the resonant frequency, respectively.

To understand the excellent filtering feature in Fig. 11, we present the intensity spectrum of the optimized cavity in Fig. 12. We can observe a resonant peak at the same frequency as in the transmission spectrum shown in Fig. 11. According to the vibration mode, most of the energy is localized in the end parts of the two solid blocks along the 45° direction. Obviously, this highly localized vibration contributes to the ultra-high *Q* factor of 6949943, as shown in Fig. 12. The simple optimized cavity combined with the waveguide in Fig. 11 may have potential applications in various small integrated acoustic or elastic wave devices.

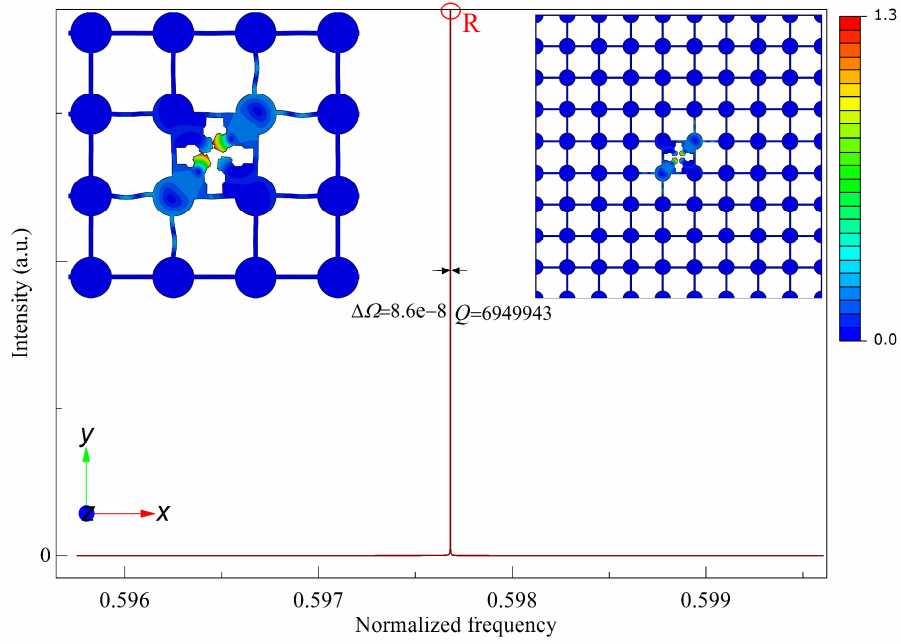

Fig. 12. Intensity spectrum of the wave transmission in the optimized cavity in Fig. 11. The insets show the spatial distribution of the amplitude of the displacement field $u=\sqrt{u_x^2+u_y^2}$ of the cavity at the resonant frequency.

After the optimized cavity is obtained, we examine the effect of the distance $H$ on the filtering property. When the cavity is moved up ($H$=0 or $a$), the filtering property at a different frequency is induced. The corresponding resonant filtering frequencies and $Q$ factors are shown in Fig. 13(a). The closer the cavity moves to the waveguide, the larger the frequency offset is, because some cavity modes are more easily to be excited by the waveguide. For example, the structure with $H=a$ has two resonant frequencies while keeping a low-$Q$ narrow drop filtering. In contrast, if the cavity is far from the waveguide (e.g., $H=3a$), the narrow connections cannot vibrate, preventing the energy flowing into the circular lumps near the cavity. For three resonant frequencies $Rf_0$, $Rf_{a1}$ and $Rf_{a2}$ in Fig. 13(a), we illustrate their responses inside the cavity in Fig. 13(b). According to the degrees of the localization, it is easy to predict that $Rf_0$ should lead to a larger $Q$ factor and a higher filtering efficiency. Similar to the mode in Fig. 12, the vibrations at the three frequencies are mainly localized on the four solid blocks in the center, but show different symmetries: symmetry for $Rf_o$, asymmetry for $Rf_{a1}$ and antisymmetry for $Rf_{a2}$. Like the structures in Fig. 8, the optimized filter in Fig. 13 can keep a certain extent of the filtering property. So, the different filtering properties in Figs. 3, 7, 8, 9 and 13 imply that the topology optimization is powerful to design wave filters under considerations of given operating frequencies, given cavity locations and/or given symmetries. This idea can be extended to realize the mode conversion [25] through the combination of the PnC cavity and the waveguide by using topology optimization in the future.

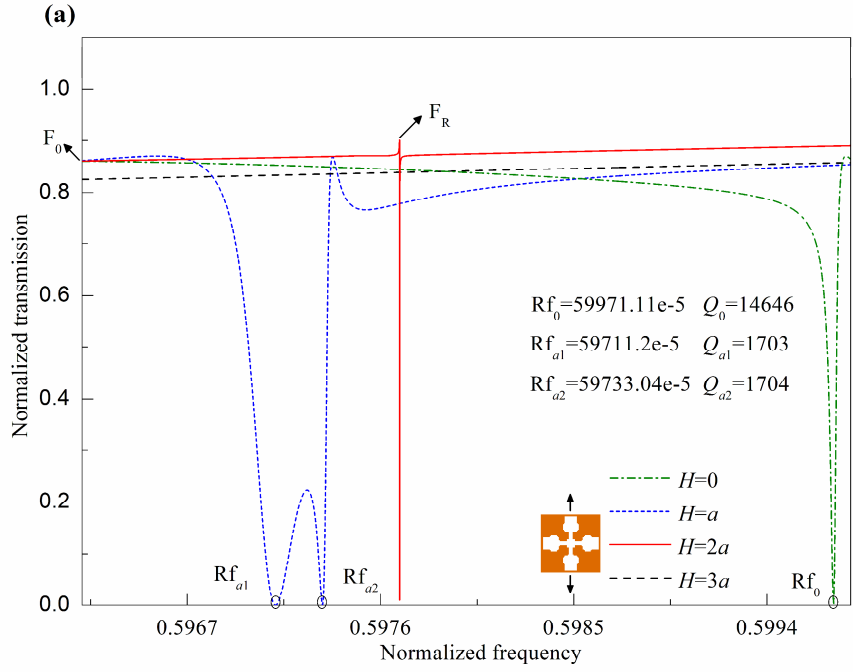

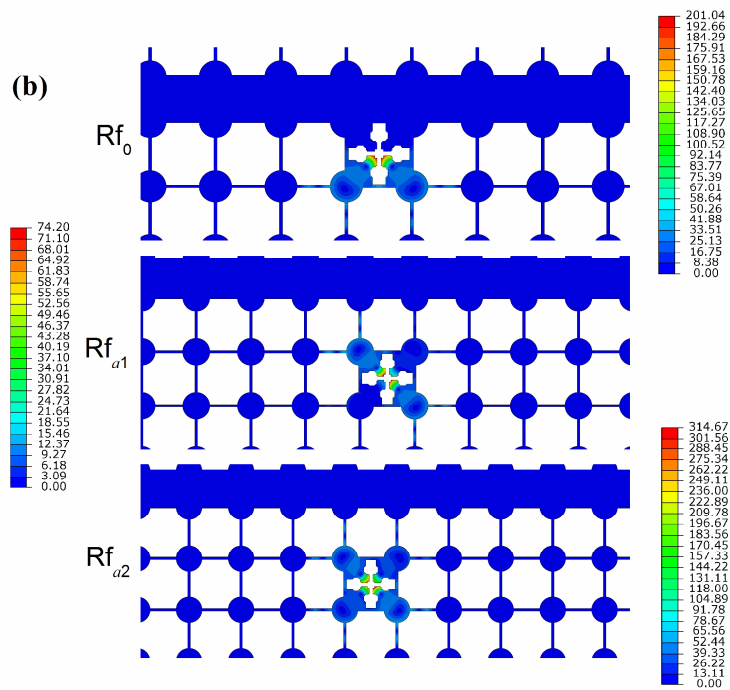

Fig. 13. **(a)** Transmission spectra of the structures by moving up ($H=0$ and $a$) and down ($H=3a$) the optimized cavity shown in Fig. 11 from the location $H=2a$. **(b)** Displacement amplitude responses inside the cavity at the three frequencies $Rf_0$, $Rf_{a1}$ and $Rf_{a2}$ marked in (a).

### (C) Fano resonance of filters

As shown in Figs. 11-13, the used local resonant unit-cell (Fig. 11) is more feasible to yield more highly localized cavity modes than the unit-cell in Fig. 1. In general, a better wave localization will lead to a lower energy loss when filtering happens. To further confirm this advantage, we perform the optimization with $H=2a$ for the incident SH wave based on the local resonant unit-cell as shown in Fig. 11. Figure 14(a) displays the normalized transmission of the optimized filter, the optimized cavity and the wave responses at three representative frequencies (A, B and C). The filtering property at the target frequency is achieved by the optimized filter. A drop efficiency of 92% and a $Q$ factor of 218442 are obtained. Obviously, in spite of the lower drop efficiency, the filter in Fig. 14(a) shows a better performance in the $Q$ factor than that in Fig. 3(b). Although we cannot prove exactly that the local resonant unit-cell is better for filtering, we can suggest that varying the based unit-cell may be helpful in the pursuit of a large $Q$ factor.

Like the results in Figs. 3, 7, 8, 9, 11 and 13, the spectrum in Fig. 14(a) shows a sharp asymmetrical line shape near the optimized resonant frequency. Here, we try to give an explanation to this unusual property. As in the case for the incident SH wave, we present the responses at the three marked frequencies A, B and C in Fig. 14(a). At the frequency without a filtering (point A), most energy is transferred into the outport from the inport. In this case, the cavity vibrates slightly because of the weak coupling. However, by comparing the responses at the points B and C, we can observe that the cavity modes are excited and most energy is flowing into the cavity. In addition, the area between the waveguide and the cavity provokes an apparent wave scattering. In other words, the energy of the resonant state lies in the energy range of the background states. And the background scattering amplitude varies slowly with the energy near the resonant energy [26]. However, the resonant scattering amplitude changes quickly in both magnitude and phase. As a result, the asymmetrical line shape is induced near the resonant frequency. As for the incident P wave, we illustrate in Fig. 14(b) the responses at the two marked frequencies $F_0$ and $F_R$ in Fig. 13(a). The enlarged view of the vibration inside the cavity at $F_R$ is shown in the bottom of Fig. 14(b). Compared with $F_0$, the cavity at $F_R$ exhibits stronger vibrations. The enlarged pictures show that they mainly vibrate at the two solid blocks along different directions. Unlike the transmission at $F_0$, the high transmission at $F_R$ is appreciably affected by the resonant cavity mode. Therefore, the interference between the background and the resonant scattering gives rise to the asymmetrical line shape. The corresponding resonant scattering phenomenon belongs to the classical Fano resonance [26, 27, 28]. This important wave phenomenon plays an important role in nonlinear wave systems, enhanced transmissions, wave switches and sensitive probes, etc. According to all Fano resonance properties of our optimized filters, we can conclude that the topology optimization is a convenient tool in designing wave filters possessing not only large $Q$ factors but also apparent Fano resonances.

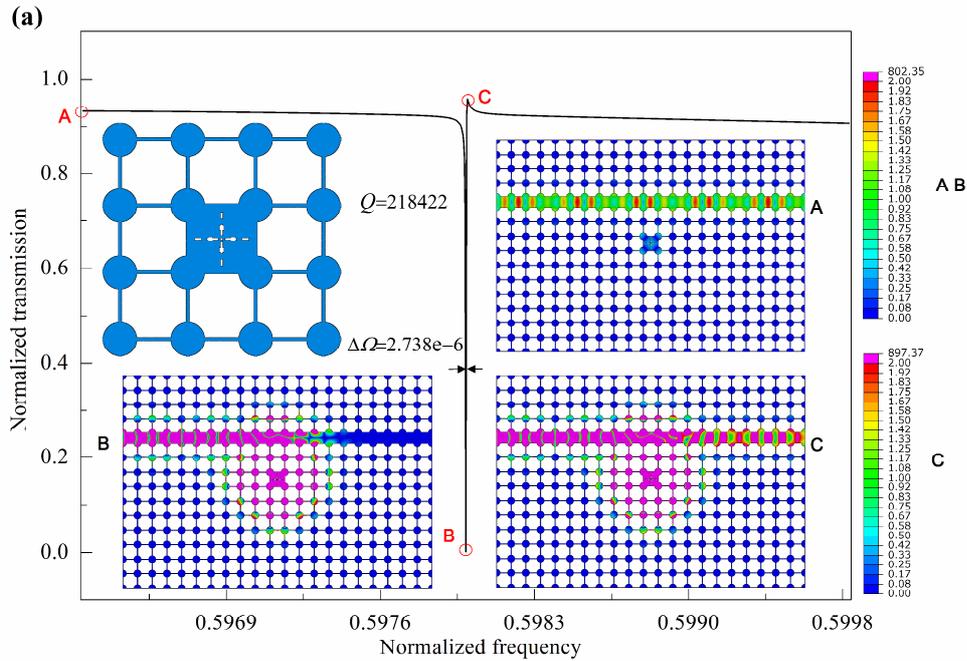

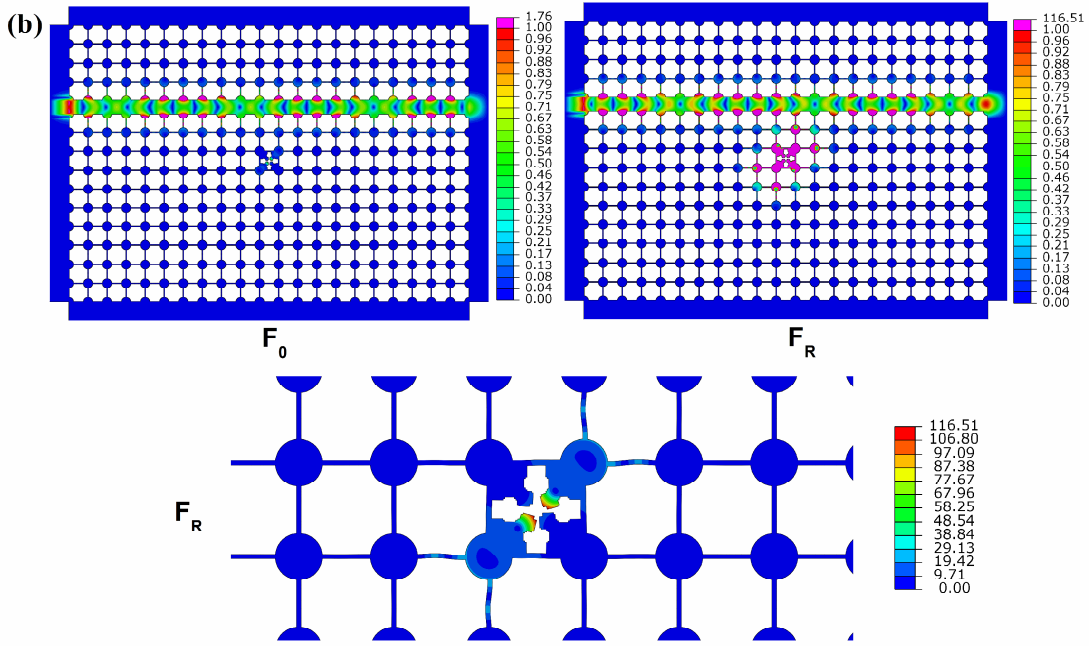

Fig. 14. **(a)** Transmission spectrum of the optimized filter for the incident SH wave at the normalized frequency 59799.98x10$^{-5}$. Responses at the three marked frequencies A, B and C are shown in the insets, respectively. **(b)** Responses at the two marked frequencies $F_0$ and $F_R$ in Fig. 13(a) and the enlarged view of the vibration inside the cavity at $F_R$.

## IV. Potential applications based on the optimized filters

### (A) Multi-wavelength filter

In view of the promising applications of high-$Q$ wave cavities and filters in PnCs, we will show how to extend the above optimized cavities of wave drop filters to other possible wave devices in this section. As shown in Figs. 3(b) and 14(a), we obtain the optimized filtering property at two different frequencies of 62631.42x10$^{-5}$ and 59799.98x10$^{-5}$, respectively. So, we combine these two filters and construct a hetero-structure as shown in Fig. 15. We consider an elastic SH wave incident from the left port (L) and right port (R) of the waveguide, respectively. It can be found here that the combined device exhibits a drop filtering at the two resonant frequencies. This important characteristic can be utilized for processing wave signals in a complex PnC system. Of course, this simple combination can be extended to filtering elastic waves of several different wavelengths or frequencies through the combination of multiple optimized filters. Besides, the combined device has the Fano resonance near the resonant frequencies as well. Since two different waveguides are connected, we argue that the difference between the two types of the transmissions with the wave incident from L or R is attributed to the reflection at the hetero-interface. On one hand, the combined device in Fig. 15 shows the possibility of operating more than one wavelength while keeping the high resolution. Furthermore, the hetero-structure shows a more flexible tunability with respect to the wavelengths. On the other hand, in contrast to changing the cavity size and the waveguide parameters, combining multiple optimized filters is easier to achieve a steady performance such as a consistent $Q$ factor and an unitary drop efficiency.

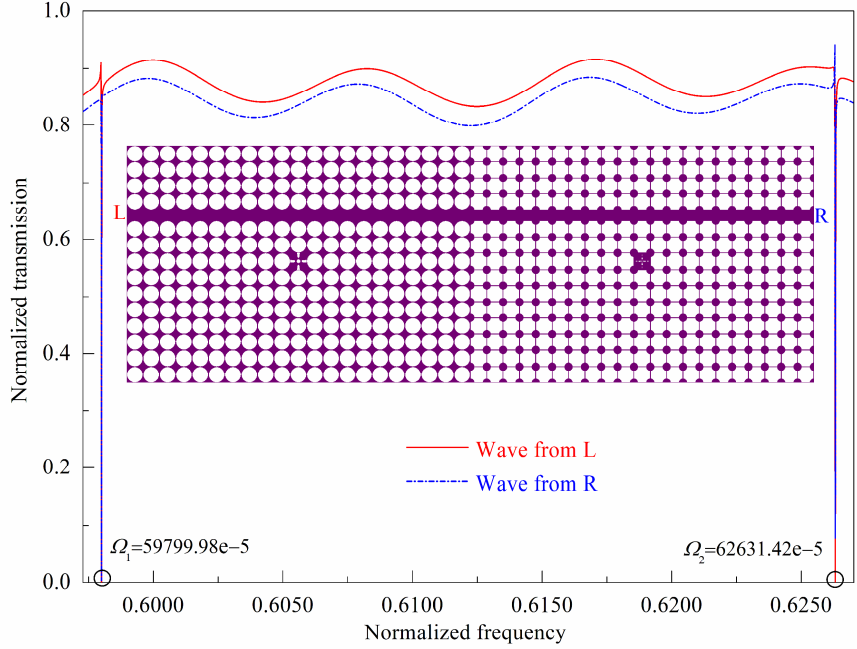

Fig. 15. Transmission spectra of the hetero-structure composed of the two optimized filters in Figs. 3(b) and 14(a). The solid and dash dot lines represent the results for the incident SH waves from the left port L and the right port R, respectively.

**(B) Raising filter**

Since the device characteristics are sensitive to any geometrical changes of the cavity, the optimized cavities with highly symmetrical modes provide a good platform for filtering elastic waves in other forms. We consider several waveguide-cavity-waveguide filters as shown in Fig. 16. The cavity in Fig. 16(a), adopted from the optimized filter in Fig. 3(b), is adjacent to two waveguides. Figure 16(b) illustrates another combination of waveguides and cavities. By changing the parameters $D$ and $W$, we present the transmission spectra of the filters and the vibration responses at the representative resonant frequencies. It is remarkable in Fig. 16(a) that the transmission property changes dramatically with the distance $D$. When the optimized defect cavity is very close to the waveguide (e.g., $D=a$), no filtering property at any frequency can be found. If $D$ becomes larger, the transmission spectrum gets narrower, resulting in a better filtering resolution. The sharp peak implies that the device acts as a raising filter. Obviously, the incident wave at the resonant frequency is less reflected than at other frequencies. For $D=3a$ and $D=4a$, their resonant frequencies $62629.81 \times 10^{-5}$ and $62629.59 \times 10^{-5}$ are nearly the same as those with $H=3a$ in Fig. 8. The high performance of the $Q$ factor (60655.1) and the raising efficiency (93.9%) demonstrates the versatility of the optimized cavity in Fig. 3(b). For the cases with two cavities in Fig. 16(b), it is possible to obtain the two resonant peaks at different frequencies. For $D=2a$ and $W=4a$, the filter has a resonant frequency of $62630.67 \times 10^{-5}$. Its $Q$ factor and filtering efficiency are 5301.4 and 81.6%, respectively. The incident wave is coupled into the upper cavity. Then, the upper cavity excites the lower cavity mode. As a result, the most energy is transferred from the upper waveguide into the lower one, see the vibration response in Fig. 16(b). Here, different values of $D$ and $W$ lead to different waveguide-cavity and cavity-cavity coupling efficiencies. These variations create the different dual-wavelength filters. Moreover, according to possible large variations of the frequencies and filtering performances, the combined filters based on our optimized cavities propose a new way to realize dual-wavelength filters. We can also introduce more optimized cavities and achieve multiple wavelength operations. The devices presented in Figs. 15 and 16(b) show some remarkable advantages of the topology optimization in a multichannel filter, i.e., a small structural size for multiple-wavelength filtering and a large

*Q* factor for large hetero-PnCs.

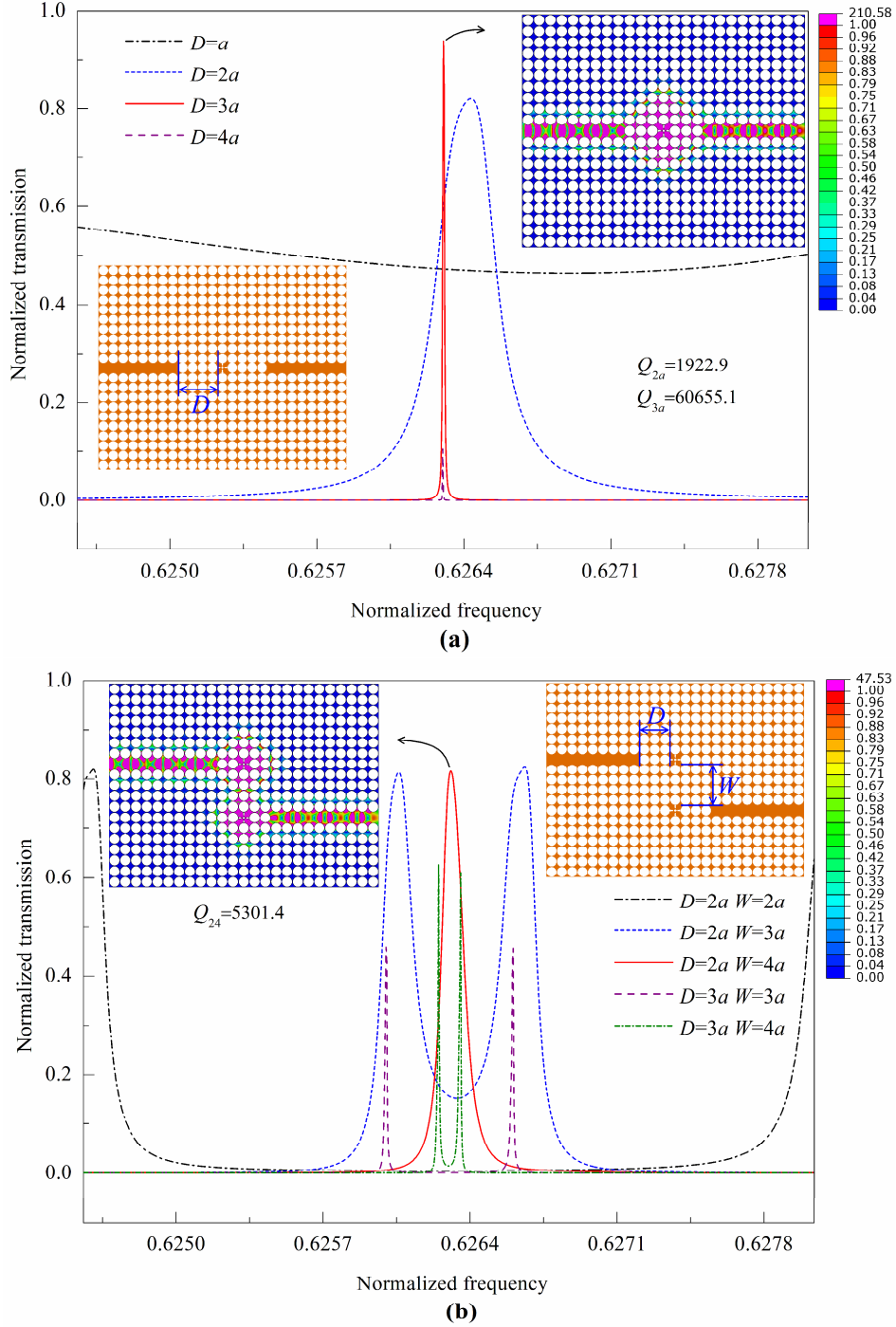

Fig. 16. Transmission spectra of the raising filter with two waveguides and one **(a)** or two **(b)** optimized cavities. The insets show the geometry of the filters and the responses at the representative resonant frequencies. The considered SH wave is incident from the left port of the waveguide.

### (C) T-splitter

More interestingly, the highly symmetrical cavity mode in Fig. 6 can be used to design a T-splitter. As a key component of a wave processing system, this waveguide device can divide the input energy equally into two output waveguides. Since the phononic bandgap can eliminate the radiation loss, we introduce the optimized cavity to reduce the reflection, see Fig. 17. Compared with the structure without a cavity, the present T-splitter with the cavity can achieve totally a 80% transmission (40% in each branch) in the frequency range from 63041.11x10$^{-5}$ to 63655.98x10$^{-5}$. The cavity successfully increases the wave transmission through the symmetrical cavity mode. Of course, a smaller *Q* factor of the cavity means a high wave transmission over a desired frequency range. However, the present T-splitter does not have a broad bandwidth. In principle, in a waveguide structure, we can insert some other medium or optimize the topology of the junction area [29] to achieve a T-splitter with nearly a 100% wave transmission, which should be a good topic to be considered in the future.

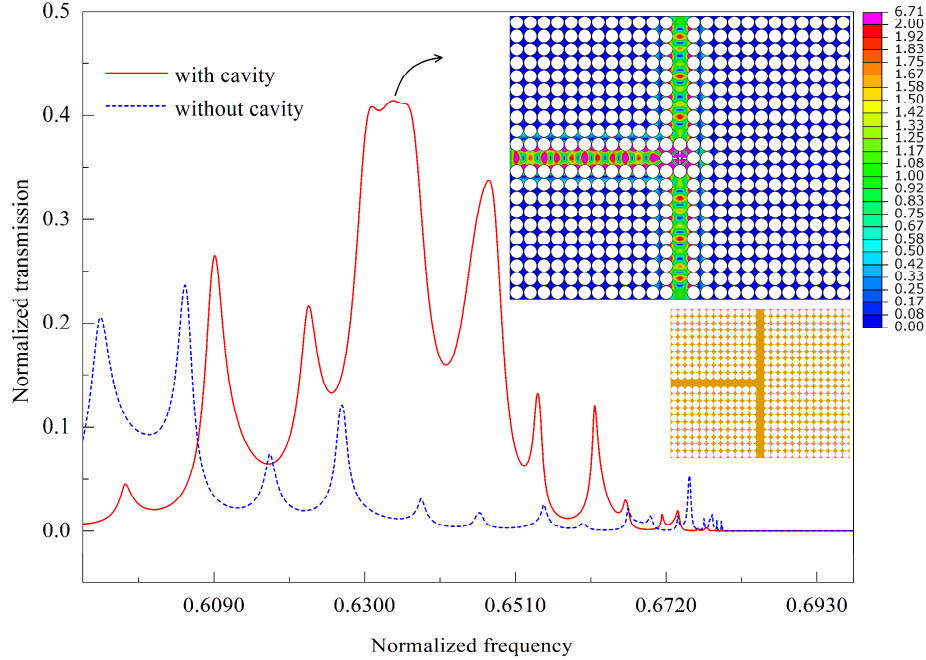

Fig. 17. Transmission spectra of a T-splitter for three waveguides with (solid line) or without (dashed line) the optimized cavity in Fig. 3(b). The insets show the vibration response of the peak transmission (upper) and the structure without cavity (lower).

## V. Conclusions

In summary, we have numerically performed the topology optimization of elastic wave filters by optimizing a cavity in 2D PnCs made by perforating holes in the elastic matrix. For the incident SH or P waves, we have investigated the effects of the target frequency of the guided wave modes and the distance between the cavity and the waveguide on the optimizations. By changing the position of the optimized cavity, we have shown how its location affects significantly the transmission property of the wave filter. Some typical wave responses at the resonant frequencies are analyzed to understand the corresponding transmission property. Furthermore, by utilizing the symmetry of the optimized cavity mode, we have further presented the applications of the optimized cavity in the multi-wavelength filters, raising filters and T-splitters, respectively. The conclusions from this investigation can be drawn as follows:

1) We can obtain a simple ultra-high-*Q* filter by only optimizing the defect cavity instead of involving the coupling medium. For the given guided wave frequencies, the optimized filters can maintain the device performance in terms of the *Q* factor and the filtering efficiency simultaneously. The symmetry of the cavity, the distance between the cavity and the waveguide and the cavity mode symmetry mainly influence the device performance. Compared with SH waves, the optimized filters for P waves are more sensitively affected by the location of the cavity. When applying the symmetrical optimized cavity in other types of the devices, it is possible to keep large *Q* factors (or high wave transmissions) and guide the elastic wave along different waveguides.

2) Most high-*Q* filters have the Fano resonances near the resonant frequencies. For the filter design, the topology optimization will inherently take out the exact coupling efficiency from the complex wave scattering. This proposes a robust way to discover more Fano resonances in PnC devices.

Indeed, other symmetries will generate different symmetrical resonant modes, resulting in the possibility to obtain an accidental degeneracy of the frequency. Therefore, in the future work, we will consider different defect types and symmetries of the cavity to design the add/drop wave filters. Without considering the uncertain structural parameters, the presented filters based on the coupled resonance will be sensitive to some specific local sizes. Therefore, it is necessary to introduce some uncertain constraints [30] to design more robust PnC devices. Evidently, we have shown that the topology optimization is well suited for the design of PnC devices by setting a simple objective function. In principle, it is ready to extend our present work to other types of PnC or PxC based devices. We can modify the objective function and then design a slow waveguide or high-*Q* cavity with a small mode volume in PnCs, a strong acousto-optic coupling cavity or waveguide in PxCs, etc. These challenging topics should be investigated in the future. In a nutshell, topology optimization offers many possibilities to design novel promising acoustic and elastic wave devices.

## Acknowledgements

This work is supported by the Fundamental Research Funds for the Central Universities (2015YJS125). The authors also acknowledge the support from the Chinese Scholarship Council (CSC) and the German Academic Exchange Service (DDAD) through the Sino-German Joint Research Program (PPP) 2014. The first author also thanks Mr. Bin Wu, Zhejiang University, PR China, for the valuable discussions and constructive suggestions.

## References

1. M. S. Kushwaha, P. Halevi, L. Dobrzynski, and B. Djafari-Rouhani, Acoustic band structure of periodic elastic composites, *Phys. Rev. Lett.* **71**(13), 2022-2025 (1933).
2. A. Khelif, A. Choujaa, S. Benchabane, B. Djafari-Rouhani, and V. Laude, Guiding and bending of acoustic waves in highly confined phononic crystal waveguides, *Appl. Phys. Lett.* **84**(22), 4400-4402 (2004).
3. M. Torres, F. R. Montero de Espinosa, J. L. Aragón, Ultrasonic wedges for elastic wave bending and splitting without requiring a full band gap, *Phys. Rev. Lett.* **86**(19), 4282-4285 (2001).
4. J. H. Sun and T. T. Wu, Propagation of surface acoustic waves through sharply bent two-dimensional phononic crystal waveguides using a finite-difference time-domain method, *Phys. Rev. B* **74**(17), 174305 (2006).
5. A. Khelif, P. A. Deymier, B. Djafari-Rouhani, J. O. Vasseur, and L. Dobrzynski, Two-dimensional phononic crystal with tunable narrow pass band: Application to a waveguide with selective frequency, *J. Appl. Phys.* **94**(3), 1308-1311 (2003).
6. C. Qiu, Z. Liu, J. Shi, and C. T. Chan, Directional acoustic source based on the resonant cavity of two-dimensional phononic crystals, *Appl. Phys. Lett.* **86**(22), 224105 (2005).
7. A. Yang, P. Li, Y. Wen, C. Yang, D. Wang, F. Zhang, and J. Zhang, High-Q cross-plate phononic crystal resonator for enhanced acoustic wave localization and energy harvesting, *Appl. Phys. Express* **8**(5), 057101 (2015).
8. Y. Pennec, B. Djafari-Rouhani, J. O. Vasseur, A. Khelif, and P. A. Deymier, Tunable filtering and demultiplexing in phononic crystals with hollow cylinders, *Phys. Rev. E* **69**(4), 046608 (2004).
9. Y. Pennec, B. Djafari-Rouhani, J. O. Vasseur, H. Larabi, A. Khelif, A. Choujaa, S. Benchane, V. Laude, Acoustic channel drop tunneling in a phononic crystal, *Appl. Phys. Lett.* **87**(26), 261912 (2005).
10. P. Zhang and A. C. To, Broadband wave filtering of bioinspired hierarchical phononic crystal, *Appl. Phys. Lett.* **102**(12), 121910

11. I. K. Lee, H. M. Seung, and Y. Y. Kim, Realization of high-performance bandpass filter by impedance-mirroring, *J. Sound. Vib.* **355**, 86-92 (2015).
12. L. Shen, Z. Ye, and S. He, Design of two-dimensional photonic crystals with large absolute band gaps using a genetic algorithm, *Phys. Rev. B* **68**(3), 035109 (2003).
13. O. Sigmund and K. Hougaard, Geometric properties of optimal photonic crystals, *Phys. Rev. Lett.* **100**(15), 153904 (2008).
14. O. Sigmund and J. S. Jensen, Systematic design of phononic band-gap materials and structures by topology optimization, *Philos. Trans. R. Soc. Lond. A* **361**(1806), 1001–1019 (2003).
15. G. A. Gazonas, D. S. Weile, R. Wildman, and A. Mohan, Genetic algorithm optimization of phononic band gap structures, *Int. J. Solids. Struct.* **43**(18), 5851–5866 (2006).
16. O. R. Bilal and M. I. Hussein, Ultrawide phononic band gap for combined in-plane and out-of-plane waves, *Phys. Rev. E* **84**(6), 065701 (2011).
17. C. J. Rupp, A. Evgrafov, K. Maute, and M. L. Dunn, Design of phononic materials/structures for surface wave devices using topology optimization, *Struct. Multidisc. Optim.* **34**(2), 111-121 (2007).
18. H. W. Dong, X. X. Su, Y. S. Wang, and C. Zhang, Topological optimization of two-dimensional phononic crystals based on the finite element method and genetic algorithm, *Struct. Multidiscip. Optim.* **50**(4), 593-604 (2014).
19. H. W. Dong, Y. S. Wang, T. X. Ma, and X. X. Su, Topology optimization of simultaneous photonic and phononic bandgaps and highly effective phoxonic cavity, *J. Opt. Soc. Am. B* **31**(12), 2946-2955 (2014).
20. J. H. Holland, Adaptation in Natural and Artificial Systems, Ann Arbor: University of Michigan Press (1975).
21. J. S. Jensen, Topology optimization problems for reflection and dissipation of elastic waves, *J. Sound. Vib.* **301**(1), 319-340 (2007).
22. O. Sigmund and J. Petersson, Numerical instabilities in topology optimization: a survey on procedures dealing with checkerboards, mesh-dependencies and local minima, *Struct. Multidisc. Optim.* **16**(1), 68-75 (1998).
23. H. W. Dong, X. X. Su, and Y. S. Wang, Multi-objective optimization of two-dimensional porous phononic crystals, *J. Phys. D: Appl Phys* **47**(15), 155302 (2014).
24. M. Notomi, K. Yamada, A. Shinya, J. Takahashi, C. Takahashi, and I. Yokohama, Extremely large group-velocity dispersion of line-defect waveguides in photonic crystal slabs, *Phys. Rev. Lett.* **87**(25), 253902 (2001).
25. G. Chen, R. Zhang, and J. Sun, On-chip optical mode conversion based on dynamic grating in photonic-phononic hybrid waveguide, *Sci. Rep.* **5**, 10346 (2015).
26. U. Fano, Effects of configuration interaction on intensities and phase shifts, *Phys. Rev.* **124**(6), 1866 (1961).
27. B. Luk'yanchuk, N. I. Zheludev, S. A. Maier, N. J. Halas, P. Nordlander, H. Giessen, and C. T. Chong, The Fano resonance in plasmonic nanostructures and metamaterials, *Nature Mater.* **9**(9), 707-715 (2010).
28. A. N. Poddubny, M. V. Rybin, M. F. Limonov, and Y. S. Kivshar, Fano interference governs wave transport in disordered systems, *Nat. Commun.* **3**, 914 (2012).
29. J. S. Jensen and O. Sigmund, Systematic design of photonic crystal structures using topology optimization: Low-loss waveguide bends, *Appl. Phys. Lett.* **84**(12), 2022-2024 (2004).
30. H. Men, K. Y. Lee, R. M. Freund, J. Peraire, and S. G. Johnson, Robust topology optimization of three-dimensional photonic-crystal band-gap structures, *Opt. Express* **22**(19), 22632-22648 (2014).